\documentstyle[mymulticol,pra,aps,amsfonts]{revtex}

\newcommand{\cW}{{\cal W}}
\newcommand{\cI}{{\cal I}}
\newcommand{\cL}{{\cal L}}

\newcommand{\cF}{{\cal F}}
\newcommand{\cB}{{\cal B}}

\newcommand{\cA}{{\cal A}}

\newcommand{\cD}{{\cal D}}

\newcommand{\beq}{\begin{equation}}
\newcommand{\eeq}{\end{equation}}
\newcommand{\beqy}{\begin{eqnarray}}
\newcommand{\eeqy}{\end{eqnarray}}

\def\C{{\mathbb{C}}}

\def\R{{\mathbb{R}}}

\def\N{{\mathbb{N}}}

\newcommand{\cH}{{\cal H}}

\title{Quantum control  without access to the controlling interaction}

\author{D. Janzing\thanks{Electronic address: janzing@ira.uka.de}, F. Armknecht, R. Zeier, and 
Th. Beth}
\address{Institut f\"ur Algorithmen und Kognitive Systeme, Am Fasanengarten 3a,
    D--76\,131 Karlsruhe, Germany}

\begin{document}
\maketitle

\begin{abstract}
In our model a fixed Hamiltonian acts on the joint Hilbert space
of a quantum system and its controller.
We show under which conditions measurements, state preparations,
and unitary implementations on the system can be performed by
quantum operations on the controller only.

It turns out that a measurement of the observable $A$  
and an
implementation of the one-parameter group $exp(iAr)$
 can be performed by 
almost the same sequence of control operations.
Furthermore  measurement procedures for
$A+B$, for $(AB+BA)$, and for $i[A,B]$ can be constructed from
measurements of $A$ and $B$. This shows that the {\it algebraic} structure
of the set of observables 
can be explained by the {\it Lie group} structure of
the  unitary evolutions on the joint Hilbert space of the measuring device
and the measured system.

A spin chain model with nearest neighborhood coupling 
shows that the border line between controller and system can be shifted
consistently. 
\end{abstract}

\begin{multicols}{2}

\section{Introduction}
Quantum state control might be described as the preparation, manipulation
and measurement of  quantum systems.
In modern research, especially in the field
of quantum optics and quantum communication,   
the control of simple quantum  systems
is one of the main goals. Typical problems are
the manipulation of 
the inner degree of an ion in a 
trap, the polarization of a photon or the preparation of Fock states
of light.
All  these manipulations of quantum systems are
performed by the interaction
between the controlled system and the particles or fields surrounding
it. Quantum state control can be formulated more precisely 
as follows: the pure states of the  quantum system
to be controlled  are described
by the  one-dimensional subspaces of a Hilbert space $\cH_s$.
For simplicity we will assume $\cH_s$ to be finite-dimensional.
Mainly, there are three actions on a quantum system which one wishes to
perform:

\begin{enumerate}
\item Initialization of a certain pure state: let $\rho$ be an arbitrary
density matrix on $\cH_s$ and $|\psi\rangle \in \cH_s$ be the vector
 of the state which should be prepared. Then initialization 
consists in causing the transition
$\rho \mapsto |\psi\rangle \langle \psi |$.

\item Unitary control: the vector state $|\psi\rangle$ is transformed
to $u|\psi\rangle$,  a density matrix $\rho$ to $u\rho u^\dagger$
for any arbitrary unitary matrix $u$.

\item Ideal quantum-non-demolition (QND) measurement:
Let $(P_i)_{i\in I}$ be a complete orthogonal family of projections
(in the sequel referred to as `the observable $(P_i)$').
Let $\cH_c$ be the Hilbert space of
any system acting as the measurement apparatus
(`c' for `controller', as measurement apparatus
and controller are in the following assumed to be given by the same system). 
A measurement of the observable $(P_i)$ is a unitary evolution
starting in a product state $|\phi\rangle\otimes |\psi\rangle \in 
\cH_c\otimes \cH_s$ and ending up in the state $\sum_i |\phi_i\rangle \otimes
 P_i|\psi\rangle$  where $(|\phi_i\rangle)_{i\in I}$ is an orthonormal family
of states of the measurement apparatus (`the pointer positions'). 
As usual, a `measurement for the observable $A$'
for any self-adjoint operator $A$ is defined by
the family of its spectral projections. 
\end{enumerate}

Initialization can be performed by a unitary control operation which is 
conditioned on the result of a former measurement. Hence we shall
only deal with measurements and unitary control operations.
Furthermore we shall not care about more general types of measurements
(positive operator valued measurements) or more general
quantum operations (completely positive maps), since they can be
obtained from
measurements and unitary operations on larger systems
(see \cite{Kr83}) by restriction to the considered system.

Common approaches to quantum control theory \cite{RSDRP95,Ll97}
are based on the assumption that the controlling person is able
to change the interaction between the system and its controller. 
This means 
that one can manipulate a (possibly time-dependent) 
Hamiltonian $H$ acting on $\cH_c\otimes \cH_s$ 
or even the system's  free Hamiltonian acting on $\cH_s$
by extern access; sometimes it is assumed that
one is able to 
control at least
a small perturbation of the Hamiltonian \cite{RSDRP95}.
Of course such an approach is well justified 
in many realistic physical situations:
macroscopic fields should rather be considered
as classical parameters determining the system's Hamiltonian
than as an interaction Hamiltonian between system and its environment.

Our motivation for investigating models
with no classical control parameters 
is twofold:

\begin{enumerate}
\item Our model can describe many physical situations where one component
of a larger
 quantum system is easier to access than the other parts:
imagine a molecule where a given control mechanism for one atom should be used
for controlling the other sort of atoms indirectly.
Another example is given by several  experiments of quantum optics:
consider an atom in a cavity  controlled by the field mode. Manipulating
the field in order to control the atom is clearly an indirect control
operation in the sense discussed below.

\item  Our model shows a symmetry between {\it measurement procedures}
and {\it unitary implementations} which is covered up in
the common approach:  
consider 
a unitary evolution $u$ on the joint Hilbert 
space $\cH_c\otimes \cH_s$ that can be decomposed into a sum 
\[
u=\sum_j v_j 
\otimes P_j
\]
 where the operators  $P_j$ are mutual orthogonal 
projections and $v_j$ are sufficiently distinct unitary transformations on 
the controller.
Then $u$ can be used for measuring an observables with spectral
projections $P_j$. By exchanging the role of controller and system one 
obtains a control mechanism for the system by switching between
 states in the image of different projections $P_j$.

For a given Hamiltonian we address the question for the set 
of possible {\it measurements} and the set of possible
{\it unitary implementations}. 
We show that the answers to both questions are given by a common theory. 
This is an important justification of our approach.
\end{enumerate}

One might argue that
our approach to the problem of quantum control
only shifts the problem to the question of how to manipulate the 
controller's quantum state.
We investigate the mathematical problems arising if such a controller
is controlled by a meta-controller and give conditions
under which this is possible.
We do not discuss the question of how to control the last controller.

Of course this problem is an extension of 
the problem
of quantum mechanical measurements (`who measures the measuring device?',
compare \cite{Neu32,He72,Em72}).
Having in mind the connection
to those  never-ending debates,
in the sequel we will be content with the result that
the borderline
between system and controller 
(so to speak the `Heisenberg cut of quantum state control')
can be shifted consistently. We are aware of the fact
that this shift causes large unsolved philosophical, physical or
biological problems, e.g.,
the problem of the interface between micro- and macrophysics\footnote{See e.g.
\cite{Pr83}.} 
(if one takes a macroscopic system as the controller), or even the 
problem of the {\it freedom of will} (the brain as the controller).
Although we ignore those problems in the sequel,
the results presented below give new insights into
the structure of the set of measurable observables and implementable
transformations.

Our approach does not refer to any assumptions on 
physical properties
of the system and the interaction. It does not even assume that the system
is a particle. It might also be one degree of freedom of a particle or
any artificial decomposition of a particle's Hilbert space to a tensor product
of two unphysical components.
The results presented below should therefore be considered as a  
structure theory of input, manipulation and read-out of quantum information.
The assumption that every dynamical evolution is caused by the Hamiltonian
represents the only physical input in such an information theory of
quantum state control.

The theory shows that the possibility of performing
every kind of measurement is equivalent to the possibility
of performing every kind of unitary transformation, since
the mathematical conditions on the form of the interaction
Hamiltonian coincide.

\section{The interface group}

We describe the quantum system 
and its controller 
by the Hilbert spaces $\cH_s$ and
$\cH_c$, respectively. For simplicity  we assume both spaces to be of finite 
dimension.
Let $W$ be the group of special 
unitary transformations acting on $\cH_c$ considered
canonically as a subgroup of the group of unitary transformations on
$\cH_c\otimes \cH_s$.
We assume 
that all that the experimentalist can do is performing
a procedure of the following form:
`perform a transformation $w_1 \in W$,
wait the time $t_1$, perform $w_2\in W$, \dots, perform $w_n\in W$, wait
the time $t_n$.' 
Formally, we denote such a procedure by
\[
p:=(t_1,w_1,\dots,t_n,w_n) \hbox{ with } t_i\geq 0, w_i\in W.
\]
We denote the set of such procedures by $P$.
The procedure $p$ implements the unitary 
\[
u_p:=w_n e^{iHt_n}\dots w_1 e^{iH t_1},
\]
where $H$ is the joint Hamiltonian on $\cH_c\otimes \cH_s$.
Its implementation time $t_p$ is 
\[
t_p:= \sum_j t_j,
\]
since we assume that the implementations on the controller
can be performed arbitrarily fast.

We consider the set
\[
I:=\overline{\{ u_p \,|\,\, p \in P\}}
\]
where $\overline{\{\}}$ denotes the closure in the norm topology.
Obviously, one has $u_p u_r=u_{rp}$ where $rp$ denotes the procedure obtained
by concatenation of the instructions $r$ and $p$.
Since the natural time evolution $e^{iHt}$
is quasiperiodic in finite dimensions, $I$ contains the unitaries $e^{iHt}$
even for negative times and therefore $I$ is a group. We shall refer to $I$ as
the {\it interface group} of the quantum control system
 $(\cH_s,\cH_c,H)$. The group $I$ describes the set of possible
transformations on the composed system.
Unfortunately, the only obvious procedure for implementing
$e^{iHt}$ for negative $t$ seems to be given by waiting for a long time until
the approximative recurrence of the system.
This can take inappropriate long times for large dimensions of $\cH_c\otimes 
\cH_s$. A priori, there is no reason why 
there should be a {\it fast} realization for $e^{-iHt}$ for small $t$. 

Here we restrict our attention to the  particular case 
where such a fast realization exists, namely
that the free Hamiltonian
of the system is zero, if the interaction Hamiltonian
is written as a sum of tensor products of traceless self-adjoint 
operators only:
the following Lemma is obtained in strong analogy to 
the first order decoupling technique in \cite{VKL98}:

\vspace{0.5cm}
\noindent
{\bf Lemma 1}
Let $H:=\sum_j A_j\otimes B_j$ be the 
 Hamiltonian
acting on the joint Hilbert space $\cH_c\otimes \cH_s$.
Assume every $A_j$ 
to be traceless.
Let $S$ be a subgroup of $W$ acting irreducibly on $\cH_c$.
Then 
\[
\sum_{s\in S} sHs^\dagger=0.
\]

\vspace{0.5cm}
\noindent
{\it Proof:} \,\,
We have $\sum sHs^\dagger= \sum_i \sum_s  sA_is^\dagger \otimes  B_i $.
Obviously, each term of the form $\sum_{s\in S} sA_is^\dagger$ 
commutes
with each $s\in S$. Due to Schur's Lemma the map 
$A_j\mapsto \sum_s sA_js^\dagger$ can only map on the 
multiples of the identity matrix. Since every $A_j$ is traceless, it  maps 
every  $A_j$ to zero.
$\Box$

\vspace{1cm}
\noindent
Hence we have $-H =\sum_{s\in S\setminus \{1\}} sHs^\dagger$.
Due to the formula
\[
\prod_{s\in S\setminus \{1\}}se^{iH\epsilon}s^\dagger 
=e^{-iH\epsilon}+O(\epsilon^2),
\] 
we  have found an approximative implementation of
the inverse time evolution $e^{-iHt}$
with the implementation time $(|S|-1)|t|$,
where $|S|$ is the group order of $S$.

In the sequel we will assume (if nothing different is said),
that  the conditions of Lemma 1 are satisfied.
This can be justified as follows:
decompose an arbitrary Hamiltonian $H$ into a sum
\[
H:=\sum_j A_j \otimes B_j + A\otimes 1 +1 \otimes B,
\]
with traceless self-adjoint operators $A_j, B_j, A, B$. 
The term $1\otimes B$ can be cancelled by passing to a rotating frame.
Apart from this we are able to cancel $A$ by an additional Hamiltonian
since we assume to have general access to the controller.
In order to work out the structure of the interface group,
we use Lie {\it algebraic} techniques:

\vspace{0.5cm}
\noindent
{\bf Definition 1} 
Let $H$ be the interaction Hamiltonian acting on the joint Hilbert space
$\cH_c\otimes \cH_s$.
Let $\cW$  be the set of self-adjoint 
traceless operators on $\cH_c$.
Then the {\it Interface Algebra} $\cI$ is the Lie algebra\footnote{We do not claim, that
$\cI$ is the Lie algebra of the Lie group $I$, since the group can be
considerably enlarged by the closure.}
generated by the  Lie sub algebra $\cW$ and the Hamiltonian $H$.

\vspace{0.5cm}

\noindent 
Then Lemma 1 allows us to
give an operational meaning to the elements of $\cI$ :

Since unitary operations of the form
$e^{\pm i H\epsilon}$ can (approximatively) be implemented during
a short time period in the order of $\epsilon$,  operations
of the form $e^{iA}$ can be implemented with arbitrarily high accuracy
if $A$ can be written as sum of multicommutators 
consisting of  $H$ and arbitrary self-adjoint operators on the controller.
Those control algorithms can be  constructed via the well-known
formulas

\[ 
e^{i(A+B)}=\lim_{m\to\infty} (e^{iA/m}e^{iB/m})^m
\]
and 
\begin{equation}\label{Kommun}
e^{-[A,B]}=\lim_{m\to\infty} (e^{iA/m}e^{iB/m}e^{-iA/m}e^{-iB/m})^{m^2}.
\end{equation}

For every $A$ in the interface algebra we can therefore approximate
every unitary of the form $e^{iAs}$ with $s\in \R$.
We rephrase this by claiming that we can simulate the unitary
time evolution corresponding to the `effective Hamiltonian' $A$.
In order to avoid misinterpretations of this suggestive 
formulation, we emphasize the following:

\begin{itemize}
\item
The simulation of the evolution $(e^{iAs})_{s\in \R}$ is discrete:
we can choose a small number $\epsilon$ and construct a control algorithm
implementing $e^{iA\epsilon}$ approximatively. By iteration, we
obtain $e^{iA\epsilon n}$ for every $n\in \N$.

\item
In general
the `simulated time' $s$ does not coincide
with the implementation time for $e^{iAs}$. Assume e.g. $A:=i[H,B]$
where $B$ is an operator on  $\cH_c$. 
Using the transformations $e^{\pm iH\epsilon}$ and $e^{\pm iB\epsilon}$
we obtain approximatively $e^{[H,B]\epsilon^2}$. This term 
is of second order in the simulation time since the latter is proportional
to $\epsilon$.
Using such a scheme, the running time for the simulation 
goes to infinity for increasing accuracy.
\end{itemize}

The interface algebra $\cI$  can be characterized explicitly:

\vspace{0.5cm}
\noindent
{\bf Theorem 1}\label{Klassifikation}
(Structure of the interface algebra)
Set $H:=\sum_j A_j\otimes B_j$ where $(A_j)$ and  $(B_j)$ are families
of  traceless 
self-adjoint
operators, linearly independent as
$\R$-vectors.
Let the dimension of $\cH_c$ be larger than $2$. Then
the interface algebra $\cI$ is given by
\begin{equation}\label{InterfaceKlass}
\cI=\cW\otimes \cB + 1\otimes \cL
\end{equation}
where $\cB$ is the self-adjoint part of
the $C^*$-algebra generated by $(B_j)$ 
and $\cL$ is the vector space  $\{i[A,B]_{A,B\in \cB}\}$.

\vspace{0.5cm}

\noindent
{\it Proof:} \,\,
Define $\cD$ as the set of self-adjoint operators $D$ such that
there is an operator $G\in \cW$ with the property
that $G\otimes D$ is in $\cI$. This is equivalent to the statement
that for every $G\in \cW$ the operator $G\otimes D$ is an element of
$\cI$ since every $\R$-linear
map on $\cW$ can be generated by concatenations and sums of maps of
the form $i[A,.]$ with $A\in \cW$ due to Lemma 2 in \cite{JB01}.
We show that $D,E \in \cD$ implies (a) $DE+ED\in \cD$ and (b)
$i[D,E] \in \cD$: take two arbitrary non-commuting 
operators $G_1,G_2 \in \cW$.
Easy calculation shows the equation
\begin{eqnarray*}
&&i[G_1\otimes D,G_2\otimes E]
+i[G_1\otimes E, G_2\otimes D]\\&=&[G_1,G_2]\otimes (DE+ ED).
\end{eqnarray*}
This proves implication (a). 

Implication (b) can be seen
as follows: choose $G_1,G_2$ such that 
$G_1^2$ and $G_2^2$ are linearly independent.
This is possible for dimension larger than $2$.
Then 
\[
i[G_j\otimes D, G_j \otimes E]=G_j^2\otimes i[D,E]  \in \cI.
\]
Hence 
\[
(G_1^2-\lambda G_2^2)\otimes i[D,E] \in \cI.
\]
Choose $\lambda\in \R$ such that $G_1^2-\lambda G_2^2$ is traceless
and hence an element of $\cW$.
This proves that $i[D,E]$ is in $\cD$.
Due to $DE=( DE+ED)/2 -(i/2)i [D,E]$ the complexification
$\cD + i\cD$ 
of $\cD$
is a algebra over $\C$ which is closed under the $(.)^*$-operation.
Hence it is a $C^*$-algebra \cite{Mu90} since it is a subset of a finite
dimensional algebra.
The following observation shows that $\cD$ contains every 
element  $B_j$:
Choose an $\R$-linear map $Y:\cW\rightarrow \cW$
with $Y(A_k)=0$ for $k\neq j$ and $Y(A_j)=A_j$.
By Lemma .. in \cite{JB01} the operator $(Y\otimes id) (H)=A_j\otimes B_j$ 
is an element of $\cI$. 
Hence $\cD+i\cD$ contains the $C^*$-algebra generated by $(B_j)_j$.  
This proves that $\cI$ contains the set $\cW\otimes \cB$.

In order to show that $\cI$ contains $1\otimes \cL$
we define $\cF$ as the set of operators $F$ such that $1\otimes F$
is an element of $\cI$.
By choosing $G\in \cW$ in such a way that $G^2=1$ one can see
that $D,E \in \cD$ implies that
\[
i[D,E] \in \cF,
\]
since $i[G\otimes D, G\otimes E]=G^2\otimes i[D,E]$. 

We have already shown  that $\cI$ contains the set
\[
\cW\otimes \cB \oplus 1\otimes \cL.
\]
Obviously this $\R$-vector space contains $\cW$ and $H$.
One checks easily that it is closed under commutators:
Let $G_1, G_2 \in \cW, A_1, A_2 \in \cB, L_1,L_2 \in \cL$. 
Then we obtain:
\begin{eqnarray*}
&&i[G_1\otimes A_1 + 1 \otimes L_1, G_2 \otimes A_2 + 1 \otimes L_2]\\&=&
i[G_1,G_2]\otimes (A_1A_2 +A_2A_1)  1/2\\&+&
(G_1G_2+G_2G_1) \otimes i[A_1,A_2] 1/2\\&+&
G_1\otimes [A_1,L_2] 
+G_2 \otimes [L_1,A_2]
+1\otimes [L_1,L_2].
\end{eqnarray*}
The terms $i[A_j,L_k]$ are clearly in $\cL$.
$A_1A_2+A_2A_1$ is an element of $\cB$.
The operator $G_1G_2+G_2G_1$ can be written as a 
linear combination of a traceless part (which is an element of 
$\cW$) and a scalar multiple of the identity.
Hence
$(G_1G_2+G_2G_1)\otimes [A_1,A_2]$ is an element of 
$\cW \otimes \cB \oplus 1 \otimes \cL$. 
Since the rhs of eq. (\ref{InterfaceKlass})
contains $H$ and $\cW$, the proof is complete.
$\Box$

\vspace{1cm}

\noindent
The assumption $dim(\cH_c)\geq 3$ can not be dropped. This could
be verified by calculating the interface algebra for
$\cH_s=\cH_c=\C^2$ with
\[
H=\sigma_x\otimes \sigma_x + \sigma_y\otimes \sigma_y,
\]
with the Pauli matrices $\sigma_x$ an $\sigma_y$.
This is 
the $x,y$-Hamiltonian (as it is often used in solid-states physics
\cite{Ar90}).
One can show  
that $\cI$ is spanned (as $\R$-vector space)
by the 10 basis vectors
\[
\sigma_x\otimes \sigma_j ,\,\, \sigma_y \otimes \sigma_j,\,\, \sigma_z \otimes 1,\,\,
1 \otimes \sigma_j
\]
with $j=x,y,z$.
We shall not bother about the case $\cH_c=\C^2$. Since we assume
to have universal access to the controller we should not only allow
unitary operations on the controller but should also take into account
unitary operations on the controller coupled to an arbitrary large ancilla 
system. Such operations might appear as non-unitary operations 
on the controller in the sense of completely positive trace preserving maps
\cite{Kr83}, measurements on the controller and so on.
Hence it makes sense to embed the original controller Hilbert space
 into a larger
controller space  $\tilde{\cH}_c:=\cH_a\otimes \cH_c$ with ancilla
space $\cH_a$ and calculate the interface algebra by taking the canonical
extension  of the original Hamiltonian $H$, namely
$\tilde{H}:=1\otimes H$.   
If we have already $dim(\cH_c)\geq 3$
an additional extension does not change the structure of the 
interface algebra given by Theorem 1
except from the fact that
$\cW$  is enlarged.
There is also another reason for extending the controller Hilbert space
in such a way: measurements are only possible if 
$\cH_c$ has sufficiently many `pointer states'.

Theorem 1 shows that universal unitary control is possible if and only
if the operators $\{B_j\}_j$ generate the full  algebra 
of linear maps on $\cH_s$.  
This can be seen as follows:
If $(B_j)$ generate the whole algebra then
$\cI$ contains  the element  $D\otimes A$ for every
self-adjoint $A$ and every traceless self-adjoint $D$. Let $\lambda_j$
be the eigenvalues of $D$ for the eigenvector $|j\rangle$.
Due
 to the equation
\[
e^{i D\otimes A s} (|j\rangle \otimes |\phi\rangle) = |j\rangle \otimes e^{i\lambda_j As} |\phi\rangle
\]
one can implement the unitary group $(e^{iAs})_{s\in \R}$.
Assume conversely that the interface algebra 
contains an operator $B$
enabling the implementation of the unitary group $e^{iAs}$ in the sense
that
 there is an initial state $|\phi\rangle$
of the controller such that 
\[
e^{iBs} (|\phi \rangle \otimes |\psi\rangle) = |\phi\rangle \otimes  e^{iAs}|\psi\rangle,
\]  
then we have 
\[
B (|\phi \rangle\otimes |\psi\rangle ) = |\phi \rangle \otimes A |\psi \rangle.
\]
Without loss of generality assume $|\phi\rangle$ to be the first
basis vector of $\cH_c$.
Define $Z:=diag(1,-1,0,\dots,0)$.
Consider the $\R$-linear map $L:\cW\rightarrow \cW$
with $L(W):= diag(\langle \psi|W|\psi\rangle, -\langle \psi|W|\psi\rangle,0,
\dots,0)$. $L$ is a (non-orthogonal) projection
on the $\R$-linear span of $Z$. Due to Lemma 2 in \cite{JB01},
$L$ can be generated by concatenations and sums of maps
of the form $i[W,.]$ with $W\in \cW$ and hence
$(L\otimes id) (B)=:C$ is in $\cI$.
Since the image of $L$ is one-dimensional, $C$ is of the form
$Z\otimes K$ with an appropriate self-adjoint operator $K$ acting on $\cH_s$.
We can show that  $K=A$ due to the equation
\[
C (|\phi \rangle \otimes |\psi\rangle) = |\phi \rangle \otimes A|\psi\rangle.
\]
Hence $Z\otimes A$ has to be in $\cI$. This implies that $A$ has to be in the  
$C^*$-algebra
generated by the set $\{B_j\}$.
Therefore we have obtained a simple necessary and sufficient condition for the
interaction
to enable universal unitary control.

\section{The `Heisenberg cut' of unitary control}

The essential result of the investigations above 
may be rephrased by the statement that
indirect quantum control is possible by operations on the controller only
(under suitable assumptions on the interaction).

The question why and in which sense  control
mechanisms for the controller  exist stands outside our theory.
At first sight, it seems as if the possibility of controlling
the controller by a meta-controller 
can obviously be explained by a mechanism of the same type.
It should be emphasized that there is 
one `detail' causing problems for doing so:
our Lie algebraic approach
assumed that the operations on the controller can be implemented arbitrarily 
fast.
This assumption can be justified if the interaction between
the meta-controller and the controller is strong compared to the
interaction between controller and system.
Recalling the fact that 
control of physical systems becomes more and more a 
matter of 
macroscopic forces if we shift the controller towards our muscles
(switching a field), this point of view might be justified.
If one does not accept such an assumption of `increasing force', 
the problem
of shifting the `Heisenberg cut of quantum control' arbitrarily
has to be  investigated more carefully. We will do this by taking 
a chain of $n$ finite dimensional quantum systems, where
each one is interacting with its 2 nearest neighbors\footnote{As an example
one might think of a spin chain as in popular models
of mathematical solid state physics \cite{Li}.}.
Than we analyze the possibilities of controlling the $n^{th}$ site
by operations on the first one.
Note that there is another detail causing difficulties in
applying the theory developed so far:
The possibility of simulating the inverse  time evolution on the
joint system $\cH_c\otimes \cH_s$ by operations on the controller only
relied on the fact that the joint Hamiltonian $H$ was assumed to have a 
decomposition into products of traceless self-adjoint operators.
This cannot be the case for a Hamiltonian
of the form
\[
H:=\sum_{j\geq 1} H_{j,j+1}
\]
where $H_{j,j+1}$ is the interaction between site $j$ and site $j+1$
since $H$ has clearly a component of the form $1\otimes A$ where
$1$ is the identity operator acting on the first site\footnote{Note that it 
would be confusing here to pass to a rotating frame with
respect to the residual Hamiltonian $\sum_{j\geq 2}H_{j,j+1}$ since
the corresponding transformation does not preserve the tensor product
structure on the remaining chain.}. 

As already mentioned, there is 
another possibility for implementing the
inverse  time evolution which is simply given by waiting.
Of course  such a technique is only realistic for quantum systems of
small dimensions.
The following theorem shows that the `Heisenberg-cut' between
the controller and the system to be controlled can be shifted arbitrarily in
in a chain of quantum systems with appropriate interactions
(provided that one accepts long waiting times for the reccurence 
of the natural time evolution):

\vspace{0.5cm}
\noindent
{\bf Theorem 2}
Let $\bigotimes_{j\leq n} \cH_j$ be the Hilbert space of a spin chain
of length $n$ with the nearest neighborhood interaction
\[
H:=\sum_{j} H_{j,j+1}.
\]
Assume every interaction term $H_{j,j+1}$ be of the form 
\[
H_{j,j+1}=\sum_k A^{(j)}_k B^{(j+1)}_k 
\]
where every $A_k^{(j)}$ is an operator acting on $\cH_j$ and
$B_k^{(j+1)}$ is an operator acting on $\cH_{j+1}$, embedded
into the space of the $n$ spins.
Let the dimension of each $\cH_j$ be greater than $2$.
For every $j$ let $\{A^{(j)}_k\}_k$ be a set of 
linearly independent self-adjoint traceless operators.
Assume that the set of operators $\{B_k{(j)}\}_{k}$ generate
the complete algebra $\cA_j$ of operators on $\cH_j$.

Then the system given by the first $m$ sites
$\bigotimes_{j\leq m}\cH_j$ can be used as a controller
for the remaining $n-m$ sites $\bigotimes_{j>m}\cH_j$, i.e., 
the Lie algebra generated by $H$ 
and the set of traceless self-adjoint operators
$(\bigotimes_{j\leq m}\cA_j)^{s.a.tl}$ is
the set of self-adjoint traceless operators  
$(\bigotimes_{j\leq n} \cA_j)^{s.a.tl}$ on the 
Hilbert space of the complete chain.

\vspace{0.5cm}

\noindent
{\it Proof:} \,\,
(By induction over $m$)
For $m=n$ the statement is trivial.
We assume it to be true for $m+1$ and conclude that it is true for
$m$:

Consider the Lie algebra $\cI_m$ generated by 
$(\bigotimes_{j\leq m}\cA_j)^{s.a.tl}$ and $H$.
Define $L$ to be the map on $\bigotimes_{j\leq m} \cA_j$ 
annihilating
every traceless operator and mapping the identity on itself.
By Lemma 2 in \cite{JB01} 
$L\otimes id^{\otimes n-m}$ can be generated by 
operations on the
controller, hence $(L\otimes id^{\otimes n-m})(H)=\sum_{m+1\leq j\leq n-1}
H_{j,j+1}$
is an element of $\cI_m$.
Since $\sum_{j\leq m-1} H_{j,j+1}$ is an element of
$(\bigotimes_{j\leq m}\cA_j)^{s.a.tl}$ we conclude that
$H_{m,m+1}$ is in $\cI_m$.
Due to Theorem 1 every Lie algebra containing
$(\bigotimes_{j\leq m}\cA_j)^{s.a.tl}$ and 
$H_{m,m+1}$ contains $\bigotimes_{j\leq m+1} \cA_j$.
This completes the induction.
$\Box$

\vspace{1cm}

\noindent
Hence we have proven that arbitrary  unitary transformations on the whole
chain can be generated by accessing the first site only.
In order to measure arbitrary observables of the chain by accessing only
the first site it is necessary to couple it to an ancilla
quantum system providing enough distinguishable quantum states. 
With the help of such an ancilla system, every measurements
can be performed: we will show in the following section
that
a controlling interaction which allows universal unitary control
allows universal measurements as well.

\section{Correspondence between measurements and implementations}

In some sense 
the problem of manipulating a quantum system by extern control mechanisms
is the inverse problem to the measurement problem. In the latter one
the state of the quantum system has an effect on the measurement apparatus's
pointer positions, in the former case the controlling apparatus has an effect
on the quantum state of the system. Whereas the philosophical problems
of understanding measurements in quantum mechanics are often discussed
in the literature, the philosophical
problems of the reverse process have hardly been discussed so far.

The symmetry between both problems can be described as follows:
a measurement can be described as unitary operations
on the apparatus  conditioned on the system state, whereas
unitary quantum state control can be described as unitary operation 
conditioned on the state (the `adjustment') of the controller.

The only asymmetry is the fact that a measurement must {\it generate} 
entanglement
for a  generic system state whereas unitary control
must {\it avoid}  entanglement with the controller.

In  our model the analogy as well as the difference  of
the  two problems  emerge  rather clearly.
We start by introducing a rather tight definition of QND-measurement
(`quantum non-demolition measurement').
Recall that a QND-measurement is usually defined 
as a measurement procedure
which does not change the eigenstates of the measured observable
and only destroys superpositions of eigenstates corresponding
to different eigenvalues.
In other words, a QND-measurement for the observable $A$ is a process  
mapping the state $|\phi \rangle \otimes |\psi \rangle$
to the state $\sum_j |\phi_j \rangle\otimes P_j |\psi\rangle$, where
$|\phi\rangle$ is the initial state of the measurement apparatus
and $|\phi_j\rangle$ are the final pointer states of the apparatus
corresponding to the spectral projections $P_j$ of $A$. They have to be
orthogonal in order to enable a perfect distinction between the different
measurement values. The notion of a QND-measurement does
not include any  assumption about the states of the  total system
{\it during} the time period $[0,T]$ between the beginning and the 
end of the procedure. In contrast, we
define a CQND-measurement ({\it continuous quantum non-demolition 
measurement}) as a measuring procedure 
where the eigenstates of the measured observable are not 
changed at  any moment {\it during} the  measurement procedure.
More precisely, we define:

\vspace{0.5cm}

\noindent
{\bf Definition 2}
An observable $A$ is said to be 
{\bf CQND-measurable} if the interface algebra $\cI$ contains
effective Hamiltonians $G_1,\dots,G_l$ and `times' $r_1,\dots,r_l$ such 
that there exists
an initial state $|\phi\rangle$ of the controller with the following
properties: 

\begin{itemize}
\item
The concatenation of the corresponding dynamical evolutions
$e^{iG_lr_l} \dots e^{iG_1r_1}$ is a measurement in the sense
explained in section 1

\item
The evolution does not disturb any eigenvector
$|\psi\rangle$ of $A$ during the measurement procedure, i.e,
for every $j\leq l$ and $r\leq r_j$ we have 
\begin{equation}\label{CQNDeq}
e^{iG_jr} e^{iG_{j-1}r_{j-1}} \dots e^{iG_1r_1}(|\phi \rangle \otimes |\psi\rangle)=|\phi_j (r)
\rangle
\otimes |\psi\rangle
\end{equation}
for some vector $|\phi_j (r)\rangle$.
\end{itemize}

\vspace{0.5cm}

\noindent
The main purpose for introducing this definition is that it 
allows the use of powerful Lie-algebraic tools in the sequel. We obtain
necessary and sufficient conditions for the existence of CQND-procedures
depending on the interface algebra. Obviously this implies necessary
conditions for existence of QND-measurements in the usual sense.

We will  show that the ability for performing CQND-measurements
of the observable $A$
is equivalent to the ability to implement the unitary one-parameter 
group $(e^{iAs})_{s\in \R}$. We define what we mean by the latter:

\vspace{0.5cm}
\noindent
{\bf Definition 3}
Let $A$ be a self-adjoint operator on $\cH_s$.
The interaction Hamiltonian $H$ is said to {\bf enable the
implementation of the group} $(e^{iAs})_{s\in \R}$ 
if the interface algebra  $\cI$ contains
an element $G$ and there is a state $|\phi\rangle$ of the controller
such that
\begin{equation}\label{implDefeq}
G (|\phi \rangle \otimes |\psi )= |\tilde{\phi}\rangle \otimes A|\psi\rangle
\end{equation}
for an appropriate vector $|\tilde{\phi}\rangle \in \cH_c$.

\vspace{0.5cm}

\noindent
Our main theorem on the set of possible implementations
and the set of possible CQND-measurements is the following:

\vspace{0.5cm}
\noindent
{\bf Theorem 3}
Let $H:=\sum_j A_j\otimes B_j$ 
be the interaction between  a quantum system and its controller
with $A_j$ and $B_j$ as in Theorem 1.
Assume  $dim(\cH_c)\geq dim(\cH_s)$.
Let $\cB$ be the $C^*$-algebra generated 
by the operators $B_j$.
If  $A$ is a self-adjoint operator acting on $\cH_s$
the following statements are equivalent:
\begin{enumerate}

\item The interaction  $H$ enables a   CQND-measurement of $A$
\item The interaction $H$ enables 
    the implementation of the one-parameter group $(e^{iAs})_{s\in \R}$
\item The operator $A$ is an element of $\cB$.
\end{enumerate}

\vspace{0.5cm}

\noindent
{\it Proof:} \,\,
$3\Rightarrow 2$: 
Due to Theorem 1 the interface algebra $\cI$ contains an
operator of the form $E\otimes A$ with arbitrary self-adjoint $E\neq 0$.
By initializing the controller to an eigenvector of $E$ 
corresponding to an eigenvalue $\lambda\neq 0$  the evolution
$e^{iE\otimes A s}$ implements $e^{iA \lambda s}$.

$3\Rightarrow 1$:
Let $(P_j)_{j\leq k}$ be the set of spectral projections of $A$.
Since $\cB$ is an algebra the operator  
$\sum_j j P_j$ is an element of $\cB$.
Due to Theorem 1 the Lie algebra $\cI$ contains
the operator $D\otimes \sum_j j P_j=\sum_j jD\otimes P_j$   
with $D:=diag(1,2,\dots,n) -(n(n+1)/2) 1$, where
$n$ is the dimension of $\cH_c$.
By initializing the controller in the state
\[
|\phi\rangle:=\frac{1}{\sqrt{n}}(1,1,\dots,1)^T
\] 
the unitary operator
\[
e^{i (\sum_j j D \otimes P_j) \pi/(2n)}
\] 
leads to mutually orthogonal final states of the controller for
the states in the image of different spectral projections $P_j$:
since $k\leq n$ (due to $dim(\cH_c)\geq dim(\cH_s)$)
the vectors $e^{ijD \pi/(2n)}|\phi\rangle$ are 
mutually orthogonal for different
$j=1,\dots,k$.

$1\Rightarrow 3$:
Choose operators $G_j$ as in Definition 2.
Define  $|\phi_j(r)\rangle$ by the equation
\[
e^{iG_jr} e^{iG_{j-1} r_{j-1}}
\dots e^{iG_1r_1}(|\phi \rangle\otimes |\psi \rangle)= 
|\phi_j (r)\rangle \otimes
|\psi\rangle , \,\,
\]
for 
 $r\in [0,r_j]$
 and every eigenstate $|\psi\rangle$ of $A$.
Define $Q_j$ to be the projection onto the span of the vectors 
$\{|\phi_j(r)\rangle \}_{r\in [0,r_j]}$.
Due to Lemma 2 in \cite{JB01} the operator
\[
\tilde{G}_j:=
(Q_j\otimes 1) G_j (Q_j\otimes 1) - tr((Q_j\otimes 1) G_j (Q_j\otimes 1))1
\] 
is an element of $\cI$.
The operator $\tilde{G}_j$ does not change any eigenstate $|\psi\rangle$
of $A$. This can be seen as follows.
\[
\tilde{G}_j|\phi_j(r)\rangle \otimes |\psi\rangle =\frac{d}{dr}
|\phi_j (r)\rangle \otimes |\psi\rangle.
\] 
Since $\tilde{G}_j$ annihilates every vector in the complement
of the span of $\{|\phi_j (r)\}$, every vector in the image
of $\tilde{G}_j$ is a product vector with $|\psi\rangle$ 
as its second component.

Therefore $\tilde{G}_j$  
can be written as a Hamiltonian on $\cH_c$ which is 
conditioned on the eigenvalue of $A$, i.e., it has the form
\[
\tilde{G}_j:=\sum_k K^{(j)}_k \otimes P_k,
\]
where $K^{(j)}_k$ are appropriate self-adjoint operators 
and $P_k$ are the spectral projections of $A$.
Since system states lying in the image of different projections $P_k$
lead to orthogonal pointer states by assumption, for every
pair $k,m$ with $k\neq m$ there is a $j$ such that
$K^{(j)}_k\neq K^{(j)}_m (mod \R 1)$.
Hence there is a linear combination
$\tilde{G}$
of the operators $\tilde{G}_j$ such that
\[
\tilde{G}=M_k \otimes P_k
\]
with $M_k \neq M_m \,\, (mod \,\,\R 1)$ for $k\neq m$.
Using the main result of \cite{JB01}
we conclude that $\cI$ contains every operators of the form
$\sum_k k D \otimes P_k=\sum_k D  \otimes kP_k $ 
with arbitrary traceless self-adjoint $D$.
Due to Theorem 1 the operator $\sum_k kP_k$ is an element
of $\cB$. Since $\cB$ is an algebra, $A$ is an element of $\cB$ too. 

$2\Rightarrow 3$: 
Define $P$ as the projection onto the span of $|\phi\rangle$ and $|\tilde{\phi}\rangle$ defined as
in equation (\ref{implDefeq}). Then $\tilde{G}:=(P\otimes 1) G 
(P\otimes 1)$ is
a nonzero operator of the form $D\otimes A$ an with appropriate operator 
$D$. Since $\tilde{G}$ is an element of the interface algebra,
the operator $A$ has to be an element of $\cB$.
$\Box$

\vspace{1cm}

\noindent
The connection between measurements and implementations
can be made even stronger in the special case that $A$ is
an observable with equidistant eigenvalues. Consider
the operator $G:=D\otimes A$ where $D$ is a operator with equidistant eigenvalues
as well. Then the same effective Hamiltonian $G$ allows 
the {\it implementation of $e^{iAs}$} by initializing the controller to an 
eigenstate
of $D$ with nonzero eigenvalue or
{\it measurements of
$A$} by taking the initial state $(1,\dots,1)^T$ for the controller.

\section{The algebraic structure of the set of measurable observables}

The constructive part of the proof of Theorem 1
gives insights to
a question which has been discussed since the early days of quantum mechanics, namely
the  algebraic structure of the set of observables.
For non-commuting observables $A$ and $B$ the spectral projections
of $A+B$ are not related to those of $A$ and $B$ in any obvious way.
Hence there is a priori no obvious connection between
measurement procedures for $A$ and $B$  and a procedure measuring $A+B$.
Similarly, there is no obvious operational meaning 
for the {\it Jordan} product $(AB+BA)/2$ and the commutator
$i[A,B]$.
P. Jordan emphasized \cite{Jo} that the equation
\[ 
AB+BA=(A+B)^2-A^2-B^2
\]
can reduce the operational meaning of the Jordan product
to the operational meaning of the addition of observables
since squaring of an observable can be interpreted as 
renaming its eigenvalues. 
He argued that the latter meaning is given by the fact
that the expectation value of $A+B$ is the sum of the expectation values
of $A$ and $B$. Since we are interested only in measurements which project
the state onto the eigenspaces, we are not content
with this explanation.
In contrast, we argue that
the {\it algebraic} structure of the measurable observables
is an implication of the {\it Lie algebraic} structure of $\cI$.
The latter structure 
has a direct operational meaning by its obvious connection
to the sequences of local operations required for performing measurements.

In order to show, that measurements of algebraic expressions
in $A$ an $B$ are related to measurements of $A$ and $B$, we are not allowed
to identify measurements of $A$ and $f(A)$ for an arbitrary
bijective function $f$ on the spectrum of $A$.
Note that we constructed measurements (in the proof of Theorem 3) 
of $A$ by simulating effective
Hamiltonians of the form $D\otimes f(A)$ where 
$D$ and $f(A)$ is an observable with equidistant eigenvalues.
In the following 
we assume a measurement procedure of any observable  $A$
to be implemented 
by an effective Hamiltonian of the form
\[
E\otimes A,
\]
with arbitrary self-adjoint $E$.
This assumption can be interpreted as follows:
If the system is in an eigenstate of $A$ corresponding to the eigenvalue 
$\lambda$ then the pointer of the measurement apparatus is moved
due to the unitary evolution corresponding to the
Hamiltonian $\lambda E$, i.e.,
the  pointer is moved in the same way having a velocity
proportional to the eigenvalue $\lambda$.
Note that (in the generic case) 
such an effective Hamiltonian will not lead to mutual orthogonal
states of the controller for different eigenvalues $\lambda$. 
There is a pragmatic solution of this problem: if the dimension
of $\cH_c$ is large there are many examples of Hamiltonians
$E$ and initial states $|\phi\rangle$  such that the states
$e^{i\lambda Es}|\phi\rangle$ are {\it almost} orthogonal 
for different eigenvalues $\lambda$.
This can be illustrated by taking an infinite dimensional space
$\cH_c:=L^2(\R)$, i.e. the set of square integrable 
functions on the real line and
think of $E$ as an operator with continuous spectrum.
Assume e.g.
\[
E:=i\frac{d}{dx}
\]  
Take small wave packages as initial states
$|\phi\rangle$ of the controller. Then 
the dynamics  
\[
e^{i\frac{d}{dx} \otimes A s}
\]
shifts the wave package
of the controller by the amount $\lambda s$  if the system's state
is an eigenstate corresponding to the eigenvalues $\lambda$.
If the width of the wave package is smaller than $s$ times the 
minimal distance of the eigenvalues of $A$, the different wave packages  
are almost orthogonal. 

Assume now we have operations on the controller
simulating the effective Hamiltonians
\begin{equation}\label{MessH}
H_A:=E\otimes A \,\,\, \hbox{ and } \,\,\,H_B:=F\otimes B
\end{equation}
with arbitrary $E,F\in \cW$.

Without knowing $A$ and $B$ itself\footnote{Note that the following
schemes assume that the experimentalist knows $E$ and $F$, but he needs
not know the Hamiltonians $H_A$ and $H_B$ completely.},
 their corresponding
 `measurement Hamiltonians'
$H_A$ and $H_B$ can be used for constructing measurements 
for
self-adjoint algebraic expressions in $A$ and $B$:

\begin{itemize}
\item {\bf measuring (A+B)}

Choose $\R$-linear maps $L_A, L_B:\cW\rightarrow \cW$ 
such that $L_A(E)=D$ and $L_B(F)=D$ with $D:=diag (1,\dots, n) -
\sum_{j\leq n} j$.
Lemma 2 in \cite{JB01} provides general rules
for simulating $\tilde{H}_A:=(L_A\otimes id)(H_A)$ and 
$\tilde{H}_B:=(L_B\otimes id)(H_B)$ 
using the time evolutions due to $H_A$ and $H_B$.
By alternating the dynamical evolution due to
the Hamiltonians $\tilde{H}_A$ and $\tilde{H}_B$ in small time steps
we obtain an approximative  simulation of
the time evolution corresponding to $D\otimes (A+B)$. If the dimension $n$ of 
the controller is  large enough, there is a parameter $s\in \R^+$
such that the unitaries $e^{i\lambda_j Ds}$ lead to 
mutual orthogonal states for different eigenvalues $\lambda_j$ of $A+B$
if the controller is initialized to the state $(1/\sqrt{n})(1,\dots,1)^T $.

\item {\bf measuring i[A,B]}

Apply small steps of time evolutions generated by
the effective Hamiltonians
$D\otimes A$ and $D\otimes B$ followed by
a simulation of the evolution due to
$-D\otimes A$ and $-D\otimes A$ (note that the linear map $D\mapsto -D$
can be implemented by operations on the controller).
This simulates an evolution corresponding to the effective Hamiltonian 
$D\otimes i[A,B]$  approximatively.
As above, this can be used for measuring $i[A,B]$.

\item {\bf measuring (AB+BA)}
 
Assume that $E$ and $F$ in eq.~(\ref{MessH}) do not commute (otherwise
one can obtain a different effective Hamiltonian
$\hat{H}_B$ by additional operations on the controller).
The equation 
\[
i[E\otimes A,F \otimes B]+i[E\otimes B, F\otimes A]=
i[E,F]\otimes (AB+BA)
\] 
provides a scheme for measuring
$AB+BA$ as follows: simulate the Hamiltonians
$\tilde{H}_B:=E\otimes B$ 
and $\tilde{H}_A:=F\otimes A$ by using operations on the controller
which exchange $E$ and $F$. 
Simulate  $i[E,F]\otimes (AB+BA)$ by evolutions corresponding to
$H_A,H_B,\tilde{H}_A,\tilde{H}_B$.
\end{itemize}

We do not expect, that the procedures
presented above are the best ones for practical purposes.
Our theory should rather illustrate  
that the algebraic  structure of the set of observables
can be understood by analyzing the continuous dynamics of the
measurement procedure.

\section{Conclusions}

We have shown that universal quantum state control does neither require
the possibility of changing the Hamiltonian of the system to be controlled
nor the ability to access its interaction Hamiltonian
to the controller. We gave necessary and sufficient
conditions  on the interaction Hamiltonian to enable universal control by
local quantum operations on the controller only.
Although the `true Hamiltonian' remains constant the experimentalists
is able to simulate arbitrary Hamiltonian time evolutions
of the controlled system. Hence our investigations
should be considered as a model for `designing' effective
Hamiltonians although we can not change the true ones.
We could show how measurement procedures for $A+B$, for $i[A,B]$
 and $AB+BA$ 
 are related to
measurement procedures of $A$ and $B$.
This gives new insights to the algebraic structure of
the set of quantum observables. We have shown that the {\it algebraic}
structure can be derived from the {\it Lie} algebraic structure
of the infinitesimal transformations on the joint system given
by the measurement apparatus and the measured system.

Furthermore we could show in which way the quantum state of a spin chain
can in principle be controlled by accessing only the first site.
This should serve as a more general model for indirect quantum control.
\end{multicols}

\section*{Acknowledgements}

Considerable improvements of the manuscript have been proposed
by P. Wocjan.
This work has partially been supported by grants of
the European Community (project Q-ACTA).

\end{document}